\documentclass[conference]{IEEEtran}

\usepackage[pdftex]{graphicx}
\DeclareGraphicsExtensions{.pdf,.jpeg,.png}

\usepackage{cite}

\usepackage[cmex10]{amsmath}
\usepackage{amsthm, amssymb}
\usepackage{graphicx}
\usepackage{xcolor}
\usepackage{booktabs}
\usepackage{mathtools}
\usepackage{multirow}

\newtheorem{proposition}{Proposition}
\newtheorem{definition}{Definition}
\newtheorem{remark}{Remark}

\newcommand{\Q}{\mathbb{Q}}

\newcommand{\Z}{\mathbb{Z}}
\newcommand{\R}{\mathbb{R}}
\newcommand{\C}{\mathbb{C}}

\newcommand{\mc}[1]{\mathcal{#1}}

\DeclareMathOperator{\mat}{Mat}
\DeclareMathOperator{\rk}{rk}

\DeclareMathOperator{\diag}{diag}
\DeclareMathOperator{\vect}{vec}

\DeclareMathOperator{\snr}{SNR}

\DeclareMathOperator{\Mat}{Mat}
\DeclareMathOperator{\ECDP}{ECDP}
\DeclareMathOperator{\Span}{span}

\begin{document}

\title{Well-Rounded Lattices for Coset Coding in MIMO Wiretap Channels}

\author{\IEEEauthorblockN{Oliver W. Gnilke, Amaro Barreal, Alex Karrila, Ha Thanh Nguyen Tran, \\ David A. Karpuk, Camilla Hollanti, \emph{Member, IEEE}}
\IEEEauthorblockA{
	Department of Mathematics and Systems Analysis\\
	Aalto University School of Science, Finland\\
	Emails: \{oliver.gnilke, amaro.barreal, alex.karrila,  ha.n.tran, david.karpuk, camilla.hollanti\}@aalto.fi}}

\maketitle

\begin{abstract}
	The concept of well-rounded lattices has recently found important applications in the setting of a fading \emph{single-input single-output} (SISO)  wiretap channel. It has been shown that, under this setup, the property of being well-rounded is critical for minimizing the eavesdropper's probability of correct decoding in lower $\snr$ regimes. The superior performance of coset codes constructed from well-rounded lattices has been illustrated in several simulations. 

	In the present article, this work is extended to fading \emph{multiple-input multiple-output} (MIMO) wiretap channels, and similar design criteria as in the SISO case are derived. Further, explicit coset codes for Rayleigh fading  MIMO wiretap channels are designed. In particular, it is shown through extensive simulations that sublattices of the well-known Alamouti code and Golden code which meet our design criteria perform better than scalar multiples of the code lattice for the same parameters.
\end{abstract}

\IEEEpeerreviewmaketitle

\section{Introduction}
\label{sec:intro}
\IEEEPARstart{I}{n} the setup of a (wireless) wiretap channel, it is assumed that the same signal is received by two different parties via two different, independent channels. The intended recipient, referred to as Bob, is assumed to have a higher quality channel and hence a higher signal-to-noise ratio ($\snr$) than the eavesdropper, Eve, who has to endure a degraded channel. This imbalance has been shown to be sufficient to achieve information theoretic security, \emph{i.e.}, Eve's received vector has negligible mutual information with the message, while there is a positive information rate from the sender, Alice, to Bob.

Code design in this setup needs to be aimed at achieving several goals simultaneously: while it is critical to maximize Bob's correct decoding probability and information rate, Eve's information needs to be minimized. Coset coding \cite{OggSoleBelfi} aims at fulfilling these goals by adding random bits to the message to confuse the eavesdropper, while allowing Bob to detect the confusion bits and correctly retrieve the information.

\subsection{Related Work and Contributions}
\label{subsec:related}

One approach to construct good lattices for coset coding in wiretap channels is to use \emph{Eve's correct decoding probability} (ECDP), studied in \cite{belfiore_rayleigh,belfiore_error}, where also design criteria were derived from approximations of the ECDP. Recently it has been shown that also in a \emph{multiple-input multiple-output} (MIMO) setting the mutual information between Eve and Alice can be related to Eve's correct decoding probability via the so-called \emph{flatness factor} of the lattice related to the eavesdropper \cite{ling, mirghasemi, luzzi}, validating this approach. For related work in terms of flatness factor and its approximations, see further \cite{karrila, barreal}.

The previous ECDP-based design criteria are commonly based on relatively coarse approximations, resulting in the so-called \emph{inverse-norm sum} in the \emph{single-input single-output} (SISO) case \cite{belfiore_rayleigh}, also studied in \cite{dave_ins}, or the \emph{inverse determinant sum} \cite{belfiore_error} in the MIMO setting. See \cite{LauraIDS} for related work in the reliability setting. In \cite{ithgain} so-called $i$-th coding gains are defined which were used in \cite{gnilke} to derive a simple geometric criterion for the design of coset codes. 

In this paper we derive a similar design criterion based on well-rounded lattices. It stems from a tighter approximation of the ECDP, and we show that it is valid for \emph{space-time} (ST) block codes. Our predictions are verified via extensive simulations and we can show that these lattices outperform the common choice of scalar multiples of the base lattice in the low SNR regime, while performing equally well for high SNR.

We introduce the required basics on lattices and cyclic algebras in Section~\ref{sec:math}, and the concept of ST coding in Section~\ref{sec:stc}, wherein we show how to construct codes from cyclic division algebras. As an example and as an ingredient for our simulations, we also introduce the famous Alamouti and Golden codes in detail. The wireless wiretap channel is covered in Section~\ref{sec:wt}, where we further introduce the concept of coset coding and derive a design criterion for MIMO wiretap coset codes. Extensive simulations are then carried out in Section~\ref{sec:sim}, where we disclose the performance of sublattices of the Alamouti and Golden codes meeting the derived design criterion when compared to other obvious choices, \emph{i.e.}, scalar multiples of the codebook lattice or other diagonal matrices.

\section{Lattices and Cyclic Algebras}
\label{sec:math}
In this section, we introduce the basic concept of a lattice and recall also basic properties of cyclic division algebras, two objects which are fundamentally important for the construction of well-performing codes for physical layer communications. 

\subsection{Lattices}
\label{subsec:lattices}
A \emph{lattice} $\Lambda \subset \R^n$ of $\emph{rank}$ $\rk(\Lambda) = s \le n$ and dimension $\dim(\Lambda) = n$ is a discrete subgroup of $\R^n$ with the property that there exist $s$ linearly independent vectors $\left(\mathbf{b}_1,\ldots,\mathbf{b}_s\right)$ of $\R^n$ such that 
	\begin{align*}
		\Lambda = \bigoplus\limits_{i=1}^{s}{\mathbf{b}_i \Z}.
	\end{align*} 
The lattice is \emph{full rank} if $s = n$.

A lattice $\Lambda' \subset \R^n$ such that $\Lambda' \subset \Lambda$ is called a \emph{sublattice} of $\Lambda$. The group \emph{index} $|\Lambda/\Lambda'|$ of $\Lambda'$ in $\Lambda$ is finite provided that $\dim(\Lambda) = \dim(\Lambda')$. 

We can conveniently represent a lattice by defining a \emph{generator matrix} $M_{\Lambda} := \left[\mathbf{b}_1\ \cdots\ \mathbf{b}_s\right] \in \mat(n\times s,\R)$, so that we can equivalently write
\begin{align*}
	\Lambda = \left\{\left.\lambda = M_\Lambda \mathbf{z} \right| \mathbf{z} \in \Z^s \right\}.
\end{align*}

The \emph{volume} of $\Lambda$ is defined to be $\nu_{\Lambda} = \left|\det(M_{\Lambda})\right|$, and is independent of the choice of basis. If $\Lambda$ is not full rank, then $\nu_{\Lambda} = \det(M_{\Lambda}^t M_{\Lambda})^{1/2}$. The volume of a sublattice $\Lambda' \subset \Lambda$ can easily been computed to be
\begin{align*}
	\nu_{\Lambda'} = \nu_{\Lambda}\left|\Lambda/\Lambda'\right|;\quad
\end{align*}

We define the \emph{Voronoi cell} associated with a lattice point $\lambda \in \Lambda$ as the set 
\begin{align*}
	\mathcal{V}_{\Lambda}(\lambda) := \left\{\left. \mathbf{x} \in \R^n \right| ||\mathbf{x}-\lambda||^2 \le ||\mathbf{x} - \lambda'||^2, \lambda' \in \Lambda \backslash \left\{\lambda\right\} \right\}.
	\end{align*}


\begin{definition}
	Let $\Lambda \subset \R^n$ be a full rank lattice, and let $\lambda_i = \lambda_{i}(\Lambda) := \inf \left\{\left. r \right| \dim(\Span(\Lambda \cap \mathcal{B}_r))\geq i \right\} $ be the succesive minima of $\Lambda$, where $\mathcal{B}_r$ is the sphere of radius $r$ around the origin. Then $\Lambda$ is called \emph{well-rounded} (WR) if $\lambda_1 = \cdots = \lambda_n$.
\end{definition}

\subsection{Cyclic Division Algebras}
\label{subsec:cda}
For a nice general exposition on cyclic division algebras and space-time codes, we refer to \cite{oggier_cda}.

 Let $L/K$ be a degree $n$ cyclic Galois field extension, and fix a generator $\sigma$ of the Galois group $\langle \sigma \rangle = \Gamma(L/K)$. A \emph{cyclic algebra} of degree $n$ is a triple 
\begin{displaymath}
	\mc{C} = (L/K, \sigma, \gamma) := \bigoplus\limits_{i=0}^{n-1}{u^iL},
\end{displaymath}
where $u^n = \gamma \in K^\times$ and $l u = u \sigma(l)$ for all $l \in L$. 
The algebra $\mc{C}$ is \emph{division}, if every nonzero element of $\mc{C}$ is invertible.

\begin{remark}
\label{rmk:quat_algebra}
	If $n = 2$, then necessarily $L = K(\sqrt{a})$ for some square-free $a \in \mathbb{Z}$. In this case, the algebra $\mc{C} = (L/K,\sigma,\gamma)$ is known as a \emph{quaternion algebra}, and can equivalently be denoted as 
\begin{align*}
	\mc{C} = (a,\gamma)_{K} \cong L \oplus j L \cong K \oplus iK \oplus jK \oplus kK, 
\end{align*}
	where the basis elements satisfy $i^2 = a$, $j^2 = \gamma$, $ij = -ji = k$. As we will see later, the case $a = \gamma = -1$ and $K =\R$ gives rise to the famous \emph{Hamiltonian quaternions} and the well-known \emph{Alamouti code}.  
\end{remark}



Given a cyclic division algebra $\mc{C} = (L/K,\sigma,\gamma)$ of degree $n$, 
the \emph{left-regular representation} is an injective algebra homomorphism $\psi: \mc{C} \to \Mat(n,\C)$ given by left multiplication $y \mapsto xy$ by a fixed $x \in \mc{C}$ for any $y \in \mc{C}$. Given an element $y = \sum_{i=0}^{n-1}{y_i u^i}\in\mc{C}$, $y_i \in \mc{O}_L$, its representation over the maximal subfield $L$ is given by 
\begin{equation}
\label{psi}
	\resizebox{0.90\hsize}{!}{$\psi: y \mapsto \begin{bmatrix} 
y_0 & \gamma\sigma(y_{n-1}) & \gamma\sigma^2(y_{n-2}) & \cdots & \gamma\sigma^{n-1}(y_1) \\
y_1 & \sigma(y_0) & \gamma\sigma^2(y_{n-1}) & \cdots & \gamma\sigma^{n-1}(y_2) \\
\vdots & \vdots & \vdots &  & \vdots \\
y_{n-2} & \sigma(y_{n-3}) & \sigma^2(y_{n-4}) & \cdots & \gamma\sigma^{n-1}(y_{n-1}) \\
y_{n-1} & \sigma(y_{n-2}) & \sigma^2(y_{n-3}) & \cdots & \sigma^{n-1}(y_0)	
	\end{bmatrix}.$}
\end{equation}

Choosing $y_i$ above to be in the ring of integers $\mc{O}_L$ will guarantee a non-vanishing determinant and hence a good coding gain, provided that the center field $K$ is either the rationals or quadratic imaginary.  This is the case for both the Alamouti code ($K=\mathbb{Q}$) and Golden code ($K=\mathbb{Q}(i), i=\sqrt{-1}$).

\section{Algebraic Space--Time Codes and MIMO channel model}
\label{sec:stc}
Algebraic \emph{space--time} (ST) coding is a powerful technique for reliable data exchange in a wireless MIMO setting. It enables spatial and temporal diversity by making use of multiple spatially separated antennas at the transmitter and~/~or receiver, and by transmitting information redundantly over multiple time instances. The well-known MIMO channel equation is given by
\begin{equation}
\label{trans_model}
	Y_{n_r\times T} = H_{n_r\times n_t}X_{n_t \times T} + N_{n_r \times T}.
\end{equation}
The subscripts $n_t$, $n_r$ and $T$ denote the number of antennas at the transmitter, receiver, and the number of channel uses, respectively. We only consider the fully symmetric case $n_t = n_r = T = n$ for some $n$, and henceforth omit all subscripts.

In the above equation, the random complex \emph{channel matrix} $H$ models Rayleigh fading, that is, the norm of its entries, $|h|$, follow a Rayleigh distribution with scale parameter $\sigma_h$, \emph{i.e.}, for every entry $h = \Re(h) + i \Im(h)$ the real and imaginary parts follow a Gaussian distribution $\Re(h), \Im(h) \sim \mc{N}(0,\sigma_h^2)$. We normalize $\sigma_h = 1$. 
The matrix $N$ is a noise matrix with zero-mean complex white Gaussian components with variance $\sigma^2$. The object of interest in the above equation is the transmitted codeword $X$, which will be an element of a finite codebook $\mathcal{X}$ of a certain algebraic structure. We assume that the channel is quasi-static, that is, $H$ stays fixed during the transmission of $X$ and then changes independently of its previous state. Perfect channel state information is only assumed at the receivers.

\begin{definition}
\label{def:stc}
	Let $\left\{ B_i \right\}_{i=1}^{k}$ be an independent set of fixed $n \times n$ complex matrices. A \emph{linear space--time block code} of rank $k$ is a set of the form
\begin{displaymath}
	\mc{X} = \left\{\left. \sum\limits_{i=1}^{k}{s_i B_i} \right| s_i \in S \right\},
\end{displaymath}
where $S \subset \mathbb{Z}$ is a finite \emph{signaling alphabet}. 

If the matrices $\left\{ B_i \right\}_{i=1}^{k}$ form a basis of a \emph{lattice} $\Lambda \subset \Mat(n, \mathbb{C})$ we call $\mathcal{X}$ a \emph{ST lattice code}. Its rank is $k = \rk(\Lambda) \leq 2n^2$, and $\mc{X}$ is full-rank in case of equality. 
\end{definition}

Henceforth, we will refer to a ST lattice code simply as a ST code. In what follows, we will quickly recall how to construct such ST codes from cyclic division algebras. The interested reader is referred to \cite{oggier_cda} for further details. An advantage of using cyclic algebras for ST coding is that a lattice structure is easily ensured by restricting the choice of elements to certain subrings of the algebra. For our examples, the restriction of the coefficients $y_i$ (cf. \eqref{psi}) to the ring of integers $\mc{O}_L$ (or an ideal therein) will suffice. This is typically referred to as the \emph{natural order} of the algebra. 

%



Let $k$ be the absolute extension degree (\emph{i.e.}, the rank of the related lattice) of $\mc{C}$ over $\mathbb{Q}$ and $\left\{ B_i \right\}_{i=1}^{k}$ a matrix basis of $\psi(\mc{C})$ over the integers $\mathbb{Z}$. A ST code constructed from the natural order for a fixed signaling alphabet $S \subset \mathbb{Z}$ is of the form
\begin{displaymath}
	\mc{X} = \left\{\left. \sum\limits_{i=1}^{k}{s_i B_i} \right| s_i \in S \right\}. 
\end{displaymath}

By choosing $\mc{C}$ to be division, we can ensure that the difference of any two distinct codewords $X-X'$ will be full-rank, and we refer to such a code as a \emph{full-diversity} code. We define 
\begin{equation*}
	\Delta_{\min}(\mc{X}) := \inf\limits_{X \in \mc{X}}|\det(X)|^2
\end{equation*} 
to be the minimum determinant of the infinite code (normalized to $\nu_\Lambda$ = 1), that is, where $S = \Z$.
As briefly mentioned earlier, the restriction of the matrix elements to (an ideal of) $\mc{O}_L$ ensures that for any matrix $\psi(y)$, $\det(\psi(y)) \in \mc{O}_{K}$, thus guaranteeing strictly positive minimum determinants $\Delta_{\min}(\mc{X})$ for $K = \mathbb{Q}$ or $K$ imaginary quadratic, even as $|S| \to \infty$.

\subsection{The Alamouti Code}
\label{subsec:alamouti}

The first ST block code was proposed in \cite{alamouti}, of which the underlying algebraic structure is that of a quaternion algebra.

Let $n = 2$. The famous \emph{Hamiltonian quaternions} can be described as
\begin{align*}
	\mathbb{H} := (-1,-1)_{\R} \cong \R \oplus i\R \oplus j\R \oplus k\R,
\end{align*}
where $\sigma$ stands for complex conjugation, and the basis elements satisfy the relation $i^2 = j^2 = k^2 = ijk = -1$.

In order to ensure a discrete structure on a ST code constructed from $\mathbb{H}$, we consider the restriction of the Hamiltonian quaternions from $\C/\R$ to $\Q(i)/\Q$, that is, consider the cyclic division algebra $\mc{C} := (\Q(i)/\Q,\sigma,-1)$, and define the \emph{Alamouti code} as a finite subset $\mc{X}_{\mathbb{H}}$ of
\begin{equation*}
	\resizebox{0.99\hsize}{!}{$\Lambda_{AC} =
	  \left\{\left. \frac{1}{\sqrt{2}} \begin{bmatrix} x_1+x_2 i & -(x_3-x_4 i) \\ x_3 + x_4 i & x_1 - x_2 i \end{bmatrix} \right| (x_1,x_2,x_3,x_4) \in \Z^4 \right\}.$}
\end{equation*}

We remark that $\mc{C}$ is a division algebra, so that $\mc{X}_{\mathbb{H}}$ is a full-diversity code. The factor $\frac{1}{\sqrt{2}}$ is in order to normalize to $\nu_{\Lambda_{AC}}=1$. Moreover, by imposing the restriction $x_i \in \Z$ on the entries of the codeword matrices, we have $\Delta_{\min}(\mc{X}_{\mathbb{H}}) = \frac14$, so that the code indeed has nonvanishing determinants.


\subsection{The Golden Code}
\label{subsec:golden}

The celebrated Golden code was introduced in \cite{golden}. For $n = 2$, consider the field extension $L/K = \mathbb{Q}(i,\sqrt{5})/\mathbb{Q}(i)$, with Galois group $\langle \sigma:\sqrt{5}\mapsto-\sqrt{5} \rangle = \Gamma\left(L/K\right)$. Define the \emph{Golden algebra} $\mc{G} = (L/K,\sigma,i)$, and let $\theta := \frac{1+\sqrt{5}}{2}$, so that $\mc{O}_L = \Z[i,\theta]$. 

Defining a ST code from $\mc{C}$ without further shaping would result in a non-orthogonal code, \emph{i.e.}, not all codeword matrices would be orthogonal. We force the code to be orthogonal by additionally considering the ideal $(\alpha) = (1-i+i\theta) \subset \mc{O}_{L}$, and define the Golden code to be a finite subset $\mc{X}_{\mc{G}}$ of
\begin{equation*}
 	\resizebox{0.99\hsize}{!}{$\Lambda_{GC} = \left\{\left.\frac{1}{5^{1/4}} \begin{bmatrix} \alpha(x_1+x_2\theta) & i\sigma(\alpha)(x_3 + x_4\sigma(\theta)) \\ \alpha(x_3 + x_4\theta) & \sigma(\alpha)(x_1+x_2\sigma(\theta)) \end{bmatrix} \right| x_i \in \Z[i] \right\}.$}
\end{equation*}

The Golden algebra is division, so that the Golden code is fully diverse. The factor $\frac{1}{5^{1/4}}$ is in order to normalize to $\nu_{\Lambda_{GC}}=1$. Moreover, it is straightforward to compute $\Delta_{\min}(\mc{X}_{\mc{G}}) = \frac{1}{5}$, so that we have nonvanishing determinants.

\section{Coset Coding for Security}
\label{sec:wt}
The wiretap channel was introduced by Wyner \cite{Wyner} and coset coding was presented as an approach to achieve secrecy in discrete memoryless channels by Ozarow and Wyner \cite{Ozarow}. Nested lattices as a realization of coset coding has been investigated in several papers for use in gaussian wireless wiretap channels, \emph{e.g.}, \cite{ ling, belfiore_error, Nested}.
When using coset coding, each message $M$ is mapped to several different codewords in $\mathcal{X}$. While this reduces the information rate, it can increase confusion at the eavesdropper. 
 
 For a given ST code $\mc{X}$ we define a coset coding scheme by fixing a sublattice $\Lambda_E \subset \mc{X}$. Two codewords $C, C' \in \mc{X}$ then represent the same message iff $C-C' \in \Lambda_E$, \emph{i.e.}, if they lie in the same coset wrt. $\Lambda_E$.
The set of all possible distinct messages $m$ 
is then given by $\mathcal{V}_{\Lambda_E}(0) \cap \mc{X}$. When sending a certain message $M$ a representative $C=M+R \in \mathcal{X}$ with $R \in \Lambda_E$ from the corresponding coset is chosen and transmitted via the MIMO channel 
\begin{equation}
Y=H (M+R)+N,
\end{equation}
as in \eqref{trans_model}.  $R$ can be either random or a public message.

As signaling alphabets we chose a $2n-PAM$ constellation $S = \{-2n+1, -2n+3, \dots, 2n-1 \}$, the odd integers in a symmetric interval around $0$. 

\begin{definition}
	The information rate in \emph{bits per ($n$) channel uses} (bpcu) $r_i=\ell_i/n$ that is achieved by this scheme is given by the number of cosets, \emph{i.e.}, the index $|\Lambda / \Lambda_E|=2^{\ell_i}$. The (average) number of coset representatives $\frac{|\mathcal{X}|}{|\Lambda / \Lambda_E|}=2^{\ell_c}$ defines the rate of confusion $r_c=\ell_c/n$ in bpcu.
\end{definition} 
We see that the total data rate is $r_d=r_i+r_c=\log_2(|\mathcal{X}|)/n$ bpcu, so that we therefore will need to balance information rate against security.

In \cite{belfiore_error}, the ECDP is used to derive a relatively complex and implicit design criterion. 
Later it was shown \cite{ling,mirghasemi, luzzi, karrila} that the ECDP is in fact related to the mutual information via the flatness factor, as mentioned in the introduction. Ignoring some constant factors it is shown that a good approximation to the ECDP is bounded by an expression of the form 
\begin{equation} \ECDP \lesssim \sum_{X \in \Lambda_E \setminus \{0\}} \det\left( I_n+ \sigma_E^{-2n} XX^* \right)^{-n_r-T} \label{ECDP}\end{equation}
where $\sigma_E^2$ is the noise variance at Eve, assuming that the fading has normalized Rayleigh parameter $\sigma_h=1$, while $n_r$ is the number of receive antennas at Eve and $T$ the number of channel uses. 
\begin{proposition}
	For low SNR the expression in equation \eqref{ECDP} is minimized by a well-rounded lattice.
\end{proposition}
A rigorous proof will be presented in an extended version of this paper, but we can quickly outline some intuitive reasoning.
We use the definition of \emph{$i^\text{th}$ normalized coding gain} from \cite{ithgain} which corresponds to the coefficient of the degree $i$ term in a polynomial expansion of the denominator in \eqref{ECDP} with respect to $\sigma_E$. The first coding gain, which determines the behaviour for low SNR is given by 
\begin{equation}
 N\delta_1(\Lambda_E):=\inf \{ \|X\|_F^2 \; | \; X \in \Lambda_E \}.
\end{equation} 
Since $\| X \|_F^2=\|\vect(X)\|_2^2$, where $\vect(X)$ is the vectorized rearrangement of $X$, we see that a design criterion for the low SNR regime should be given by maximizing the minimum length of vectors in $\Lambda_E$. From Minkowski's second theorem we know that for lattices of fixed volume the product of all successive minima is bounded. The minimum is therefore maximized in the case where all succesive minima are equal, leading us to consider WR lattices as choices for $\Lambda_E$, similarly to the SISO case \cite{gnilke}.

\section{Simulations}
\label{sec:sim}
\subsection{Alamouti}
We compare several sublattices of the Alamouti code using the signaling alphabets $S_1:=\{\pm 3, \pm 1\}$ or $S_2:=\{\pm 7, \pm 5, \pm 3, \pm 1\}$ which correspond to a 4-PAM (equivalently, 16-QAM) or an 8-PAM (equivalently, 64-QAM) constellation for the real (equivalently, complex) symbols, respectively. The Alamouti code as a sublattice of $\R^8$ is described by the generator matrix 

\begin{equation*}
A:= \frac{1}{\sqrt{2}}\begin{bmatrix}
1  &   0  &   0   &  0\\
0  &   1  &   0   & 0\\
0  &   0  &   1   &  0\\
0  &   0  &   0   &  1\\
0  &   0  &  -1  &   0\\
0  &   0  &   0   &  1\\
1  &   0  &   0   &  0\\
0  &  -1 &    0   &  0\\
\end{bmatrix}. 
\end{equation*}
The codebook is then given by all vectors in $\mathcal{X}=\{A\mathbf{z} : \mathbf{z} \in S^4\}$. For our simulations, we fix $n_r = 2$ and proceed in the following fashion. After choosing a random element $X = Ax \in \mc{X}$, we calculate $Y$ as in equation \eqref{trans_model} and use a sphere decoder to find the closest vector $Z \in \mc{X}$ in the codebook. Since we apply coset coding we have to consider the case that $Z \neq X$, but $X \equiv Z \mod \mc{X}_E$, where $\mc{X}_E \subset \mc{X}$ is a subset, in which case $X$ and $Z$ represent the same message. Instead of defining $\mc{X}_E$ directly as a sublattice of the Alamouti code we define a sublattice of the coefficient set $\Z^4$, \emph{i.e.}, let $X = Ax$ and $Z = Az$ then $X \equiv Z \mod \mc{X}_E \; \Leftrightarrow \; x \equiv z \mod \Lambda_E$. Since $A$ is orthonormal, a well-rounded sublattice $\Lambda_E \subset \Z^4$ will define a well-rounded sublattice $\mc{X}_E \subset \mc{X}$. 

We define three different such sublattices $\Lambda_i$ by their generator matrices $L_i$. Each of these lattices provides the same number of cosets and hence the same information rate. The first lattice is a straightforward approach of achieving index $32$ in the signaling set, simply by constructing a diagonal matrix. The second matrix has been found by simple computer search and provides the same index, while having a larger minimal length. The lattice $\Lambda_3$ is a scaled version of the $D_4$ lattice. The fourth and fifth lattices have index $256$ and are a simple multiple of the base lattice and an optimized WR lattice found by computer search.


\begin{align*}
L_1 &:= 2 \cdot \begin{bmatrix} 4  & 0  &   0   &  0 \\ 0  &   2  &   0   & 0 \\ 0  &  0  &  2   &  0 \\ 0  &  0  &   0   &  2 \end{bmatrix}, \;\; L_2 := 2 \cdot \begin{bmatrix} -2  &   -2  &   0   &  0 \\ 0  &   0  &   -2   & -1 \\ -1  &   1  &   1   &  -2 \\ 1  &   -1  &   1   &  -1 \end{bmatrix} \\
L_3 &:= 2 \cdot \begin{bmatrix}4  &   2  &   2   &  2\\0  &   2  &   0   & 0\\0  &   0  &   2   &  0\\0  &   0  &   0   &  2\end{bmatrix}, \;\; L_4 := 2\cdot \begin{bmatrix} 4  & 0  &   0   &  0 \\ 0  &   4  &   0   & 0 \\ 0  &  0  &  4  &  0 \\ 0  &  0  &   0   &  4 \end{bmatrix} \\
L_5 &:= 2 \cdot \begin{bmatrix} -2 & -3& 4& -1\\0 &-1& 0& 3\\0 &-3& -2& -3\\-4 &-1& 0& -1\end{bmatrix}. 
\end{align*}

\begin{figure}
		\centering
	\includegraphics[scale=0.44]{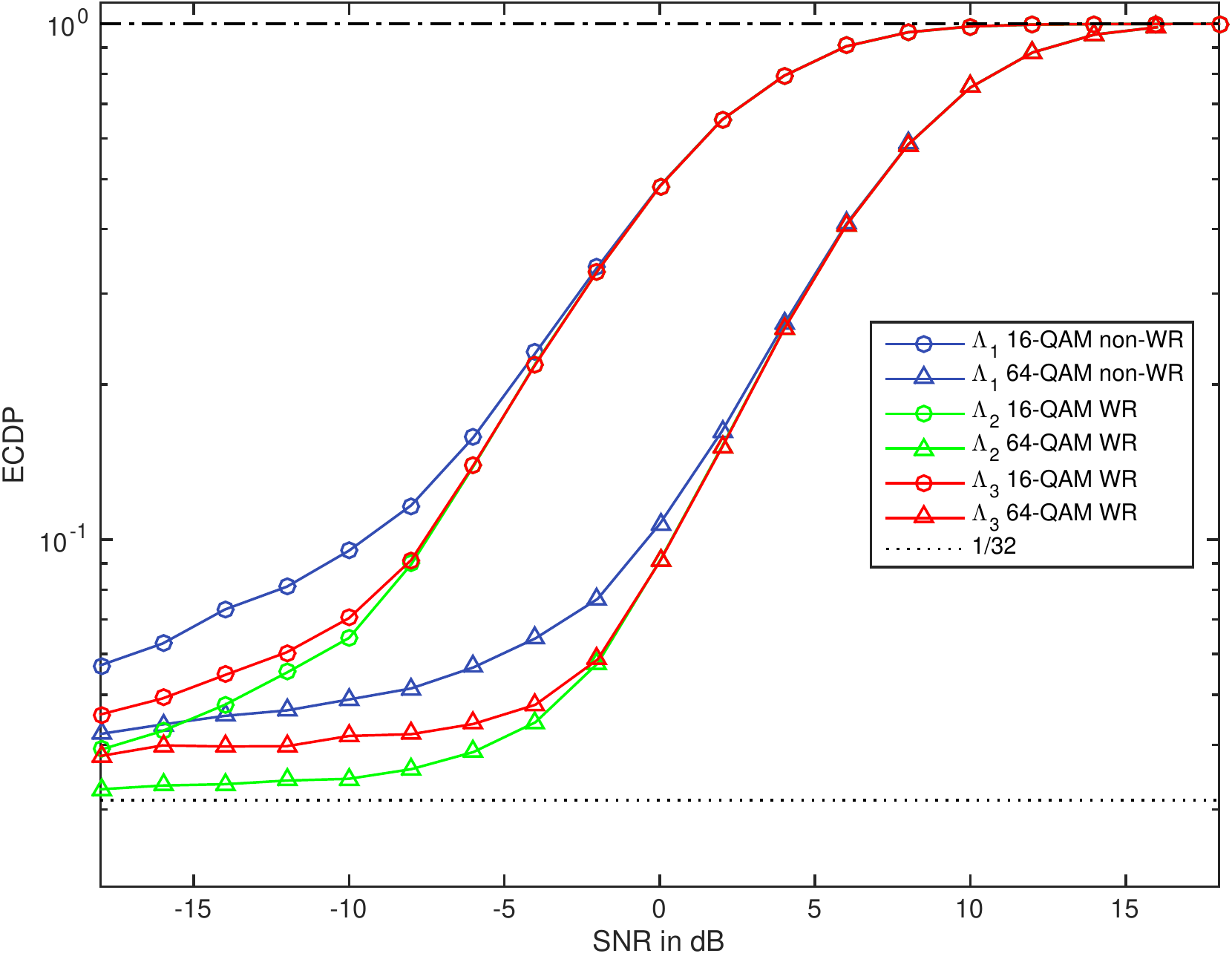}
	\caption{ECDP for 16-QAM and 64-QAM Alamouti}
	\label{Alamouti_fig}
\end{figure}

The two different signaling sets $S_1$ and $S_2$ give us two different codebook sizes. Since the information rate is fixed by the index of the sublattice, an increase in codebook size increases the rate of confusion. The ECDP is lower bounded by the reciprocal of the index, since a correct decoding rate of $\frac{1}{32}$ is equal to randomly guessing the message, hence no mutual information between Alice and Eve. In Figure \ref{Alamouti_fig} we can see that the different lattices perform equally well in the high SNR regime, providing the legitimate receiver with comparable reliability. For lower SNRs the different lattices show different performances and it can be seen that the WR lattices approach the lower bound quicker than the non-WR lattices. The higher rate of confusion in the bigger codebooks translates into a $10$dB gain.

\begin{table}[h!]
	\begin{center}
	\begin{tabular}{c r r r | r r r | r r r}
		\toprule
			&			&			&							& \multicolumn{3}{c|}{16-QAM} & \multicolumn{3}{c}{64-QAM}\\
			& index	&	WR	&	$\lambda_1^2$ & $r$ & $r_i$ & $r_c$ & $r$ & $r_i$ & $r_c$ \\ 
		\midrule
			
		$\Lambda_1$& 32 & no & 16 &  & & & \\
		$\Lambda_2$& 32 & yes & 24 & 4 & 2.5& 1.5& 6 & 2.5 & 3.5\\
		$\Lambda_3$& 32 & yes & 32 &  & & &\\
		\midrule
		$\Lambda_4$& 256 & yes & 64 & & & & \multirow{2}{*}{6} & \multirow{2}{*}{4} & \multirow{2}{*}{2}\\
		$\Lambda_5$& 256 & yes & 80 & & & &  &  & \\
		\bottomrule\\
		\end{tabular}
	\caption{Sublattices of the Alamouti Code}	\label{sublatAlam}
	\end{center}
\end{table}

\begin{figure}
		\centering
	\includegraphics[scale=0.44]{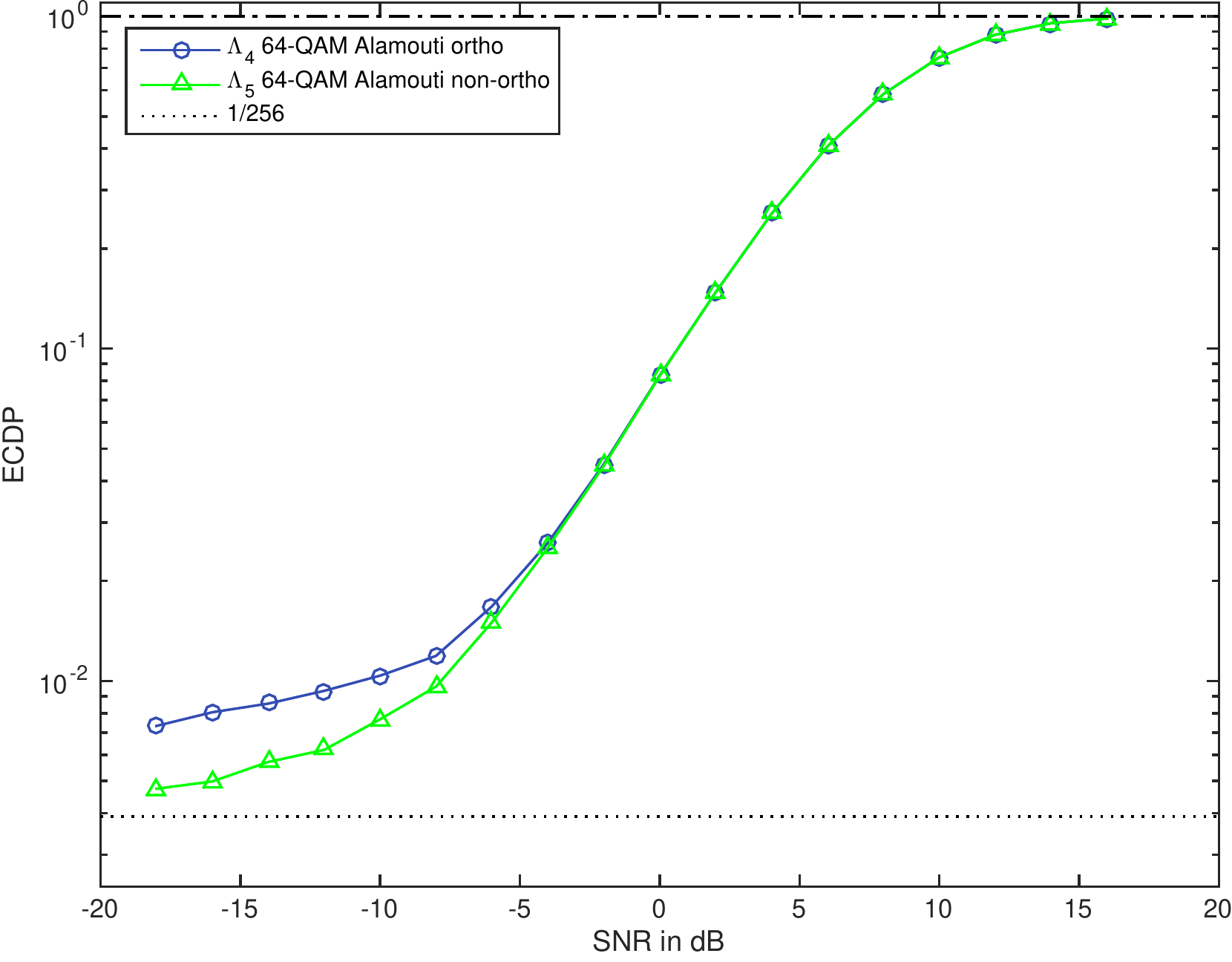}
	\caption{ECDP for 64-QAM Alamouti}
	\label{Skew_fig}
\end{figure}

Identical to the observations in \cite{skewed, gnilke} we see in Figure \ref{Skew_fig} that it is advantageous to use WR lattices that are not orthogonal. 

\subsection{Golden Code}
The Golden code is an orthonormal sublattice of $\R^8$ with generator matrix
\begin{equation*}
\resizebox{0.99\hsize}{!}{$G:= \frac{1}{\sqrt{5}}\begin{bmatrix}
1&1-\theta&\theta&-1&0&0&0&0\\
\theta-1&1&1&\theta&0&0&0&0\\
0&0&0&0&1&1-\theta&\theta&-1\\
0&0&0&0&\theta-1&1&1&\theta\\
0&0&0&0&1-\bar{\theta}&-1&-1&-\bar{\theta}\\
0&0&0&0&1&1-\bar{\theta}&\bar{\theta}&-1\\
1&1-\bar{\theta}&\bar{\theta}&-1&0&0&0&0 \\
\bar{\theta}-1&1&1&\bar{\theta}&0&0&0&0\\
\end{bmatrix}$}
\end{equation*}
where $\bar{\theta}=\frac{1-\sqrt{5}}{2}$. As in the previous subsection we choose codebooks by defining finite signaling sets. We again use odd integers in a finite range which can be understood as QAM constellations, \emph{e.g.}, $S_1:=\{ \pm 1 \}$ corresponds to $4$-QAM, $S_2:=\{\pm 3, \pm 1 \}$ corresponds to $16$-QAM, \dots, $S_4:=\{\pm 9, \pm 7,\dots, \pm 1\}$ corresponds to $100$-QAM.

We chose sublattices $\Lambda'_i$ of index $32$ with generator matrices
\begin{align*} 
L'_1 &= 2\cdot \diag(2,2,2,2,2,1,1,1),\\
L'_2 &= 2\cdot\begin{bsmallmatrix}
 1& 0& 1& 0& 0& 0& 1& 1\\
 0& 0& 0& 0& 1& -1& 0& -1\\
 1& 1& 0& 0& 0& -1& -1& 0\\
 0& -1& -1& 1& 0& 0& 0& 0\\
 0& 0& 0& 1& 0& 1& 0& -1\\
 -1& 0& 0& 1& 0& 0& 0& 0\\
 0& 1& -1& 0& 1& 0& 0& 0\\
 0& 0& 0& 0& -1& 0& 1& 0\\
\end{bsmallmatrix},\\
L'_3 &= 2\cdot \begin{bsmallmatrix} 
 -1& -1& -1& 1& 0& 0& 0& 0\\
 1& 0& 0& 1& 0& 0& 0& 0\\
 0& 0& 1& 1& 0& 0& 0& 0\\
 0& -1& 0& 0& -2& 0& 0& 0\\
 -1& -1& 0& 0& 0& 2& 0& 0\\
 -1& 0& 1& 0& 0& 0& 0& 2\\
 0& 1& 0& -1& 0& 0& 0& 0\\
 0& 0& -1& 0& 0& 0& -2& 0\\
\end{bsmallmatrix}.
\end{align*}

The first lattice is again a simple diagonal construction, while the other two were found by computer search.
\begin{table}[h!]
	\setlength{\tabcolsep}{4pt}
	\begin{center}
		\begin{tabular}{c r r r | r r r | r r r | r r r}
			\toprule
			&			&			&							& \multicolumn{3}{c|}{4-QAM} & \multicolumn{3}{c}{16-QAM} & \multicolumn{3}{c}{64-QAM}\\
			& index	&	WR	&	$\lambda_1^2$ & $r$ & $r_i$ & $r_c$ & $r$ & $r_i$ & $r_c$ & $r$ & $r_i$ & $r_c$\\ 
			\midrule
			
			$\Lambda'_1$& 32 & no & 4 &  & & & & & & \\
			$\Lambda'_2$& 32 & yes & 12 & 4 & 2.5& 1.5& 8 & 2.5 & 5.5 & 12 & 2.5 & 9.5 \\
			$\Lambda'_3$& 32 & yes & 16 &  & & & & & &\\
			\bottomrule\\
		\end{tabular}
		\caption{Sublattices of the Golden Code of index $32$}	\label{sublatGold}
	\end{center}
\end{table}

\begin{figure}
		\centering
	\includegraphics[scale=0.44]{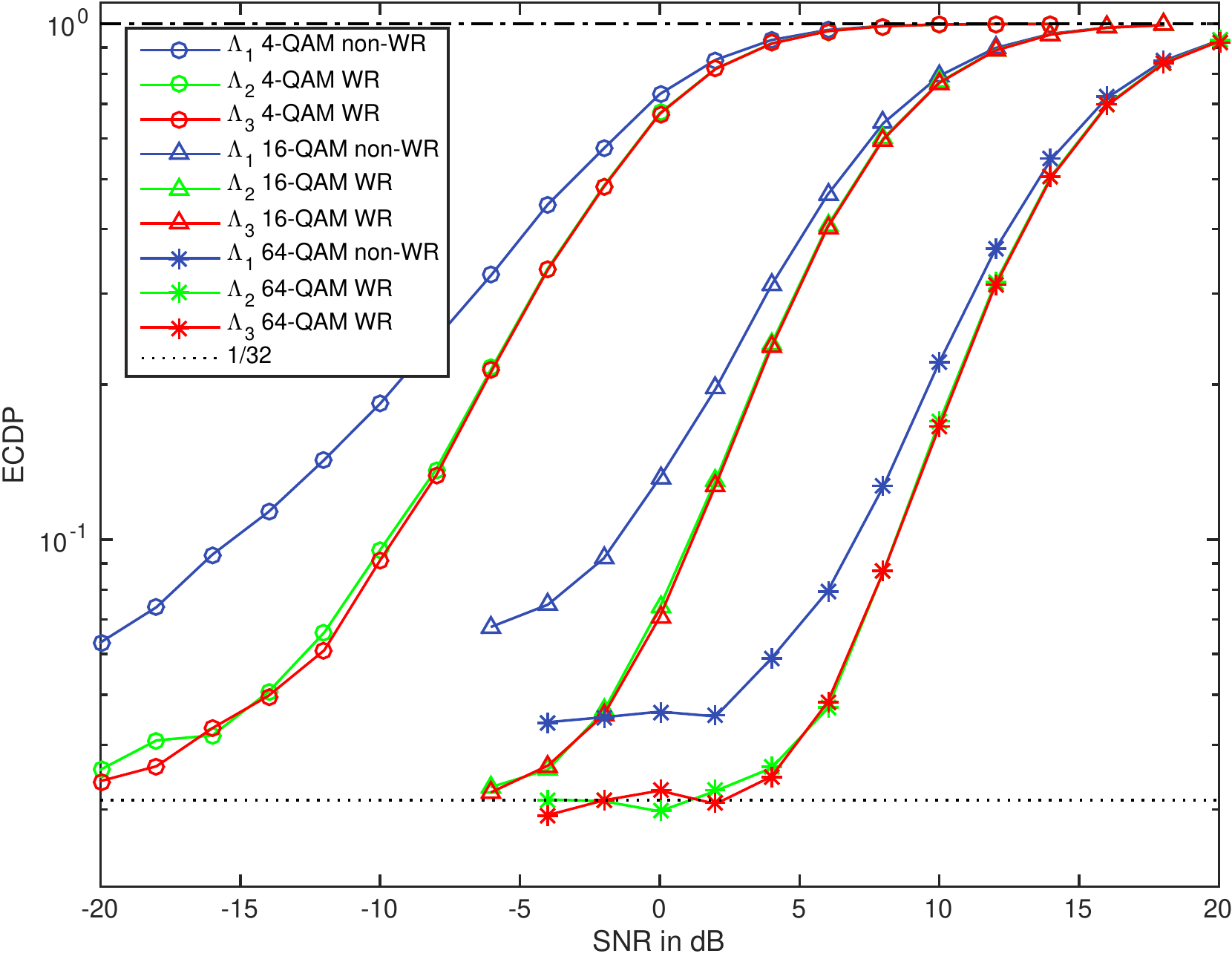}
	\caption{ECDP for 4-QAM, 16-QAM and 64-QAM Golden Code}
	\label{Gold1_fig}
\end{figure}

The simulation results in Figure \ref{Gold1_fig} show even more pronouncedly than in the Alamouti case that the WR lattices outperform the non-WR choices, while maintaining reliability in the high SNR regime. Also the effect of adding more bits of confusion is clearly visible. Again there is a 10dB gap between 1.5 and 5.5 bits of confusion, and a 8dB difference from 5.5 to 9 bits of confusion.

\begin{figure}
		\centering
	\includegraphics[scale=0.42]{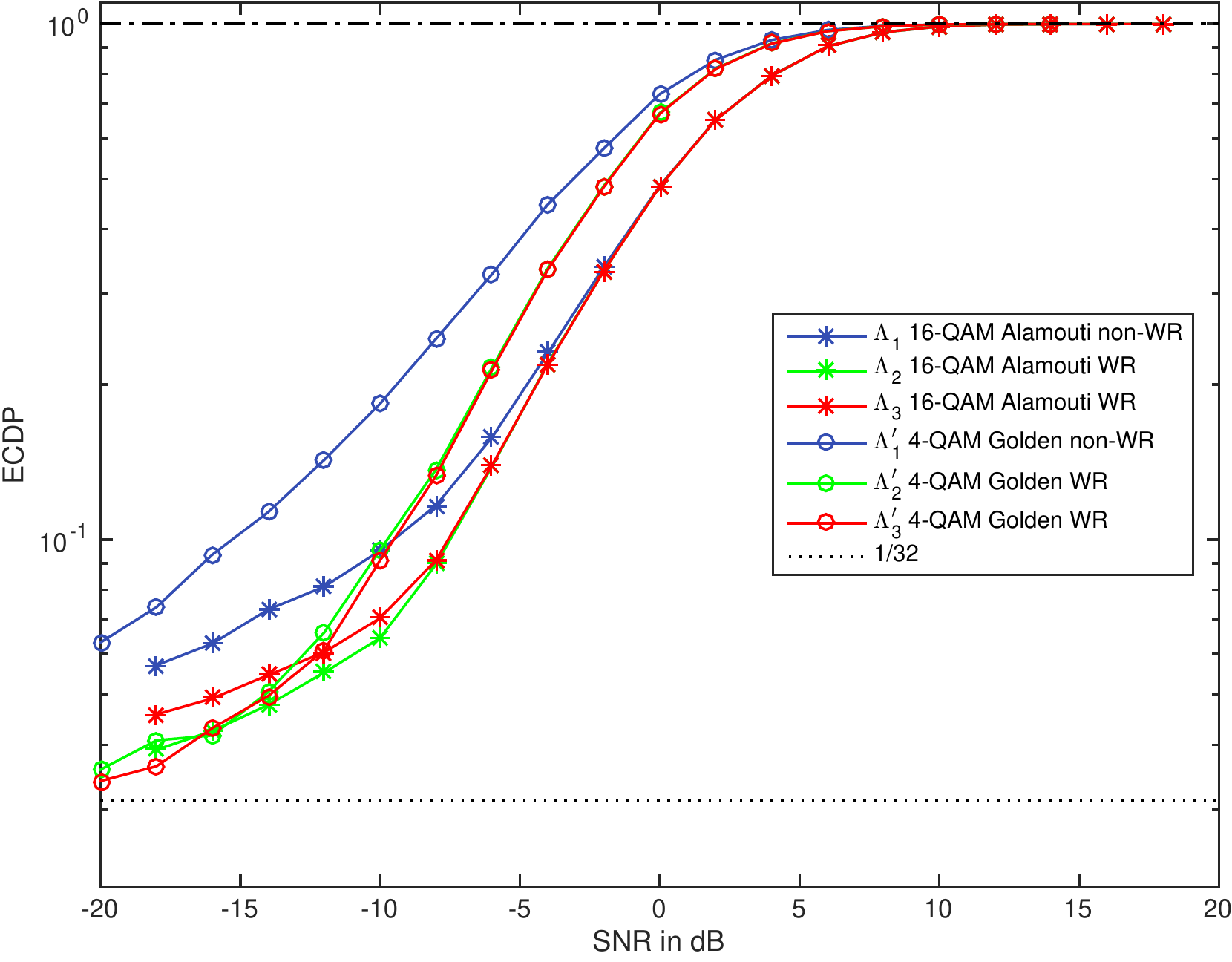}
	\caption{ECDP for 16-QAM Alamouti and 4-QAM Golden Code}
	\label{Comparison_fig}
\end{figure}

A comparison between the Alamouti and the Golden code in Figure \ref{Comparison_fig} shows that the Golden code achieves higher reliability in the high SNR regime, but catches up with even the best sublattice of the Alamouti code in low SNR.
To investigate the effect of the ratio between $r_i$ and $r_c$ we run a simulation with three sublattices that have comparable values $r_c$ but different $r_i$, when combined with the right signaling set. Their generator matrices are given by
\begin{align*} 
M_1 &= 2\cdot\begin{bsmallmatrix}
0& 2& 0& 0& 0& 0& -1& 0\\ 
0& 0& 0& 0& 0& 1& -1& 0\\ 
0& 0& 0& 0& -2& 0& 0& -1\\
0& 0& 2& 0& 0& 0& 1& 0\\ 
0& 0& 0& 2& 0& 1& 0& 0\\ 
-2& 0& 0& 0& 0& 1& 0& 1\\
0& 0& 0& 0& 0& 0& 1& -1\\
0& 0& 0& 0& 0& 1& 0& 1\\
\end{bsmallmatrix}\\
M_2 &= 2\cdot\begin{bsmallmatrix}
-1& 0& 0& 2& 0& 0& -4& 0\\ 
2& 1& 0& 0& -4& 2& 0& 0\\ 
1& 2& 0& 1& 0& -3& 0& 0\\ 
0& -1& 0& 2& 0& -1& 0& 0\\
0& 1& 0& 1& 0& 0& 0& 4\\ 
3& 0& 0& -1& 0& 1& 0& 0\\
0& 3& 0& 1& 0& 0& 0& 0\\
-1& 0& -4& -2& 0& 1& 0& 0\\
\end{bsmallmatrix},\\
M_3 &= 2\cdot \begin{bsmallmatrix} 
-2& -2& -2& 0& -4& -2& 0& 0\\
0& 0& 2& 0& 0& -1& 0& 4\\ 
0& 1& -2& 1& 0& 2& 0& 1\\ 
-1& 0& -2& 0& 0& -2& 5& 0\\ 
-4& 0& 1& -2& 0& 0& 0& -3\\ 
-2& -1& 0& 4& -1& 2& 0& 0\\ 
1& -4& 0& 1& 3& 0& 1& 0\\ 
0& -2& -3& 2& 0& -3& 0& 0\\
\end{bsmallmatrix}.
\end{align*}

\begin{figure}
	\centering
	\includegraphics[scale=0.42]{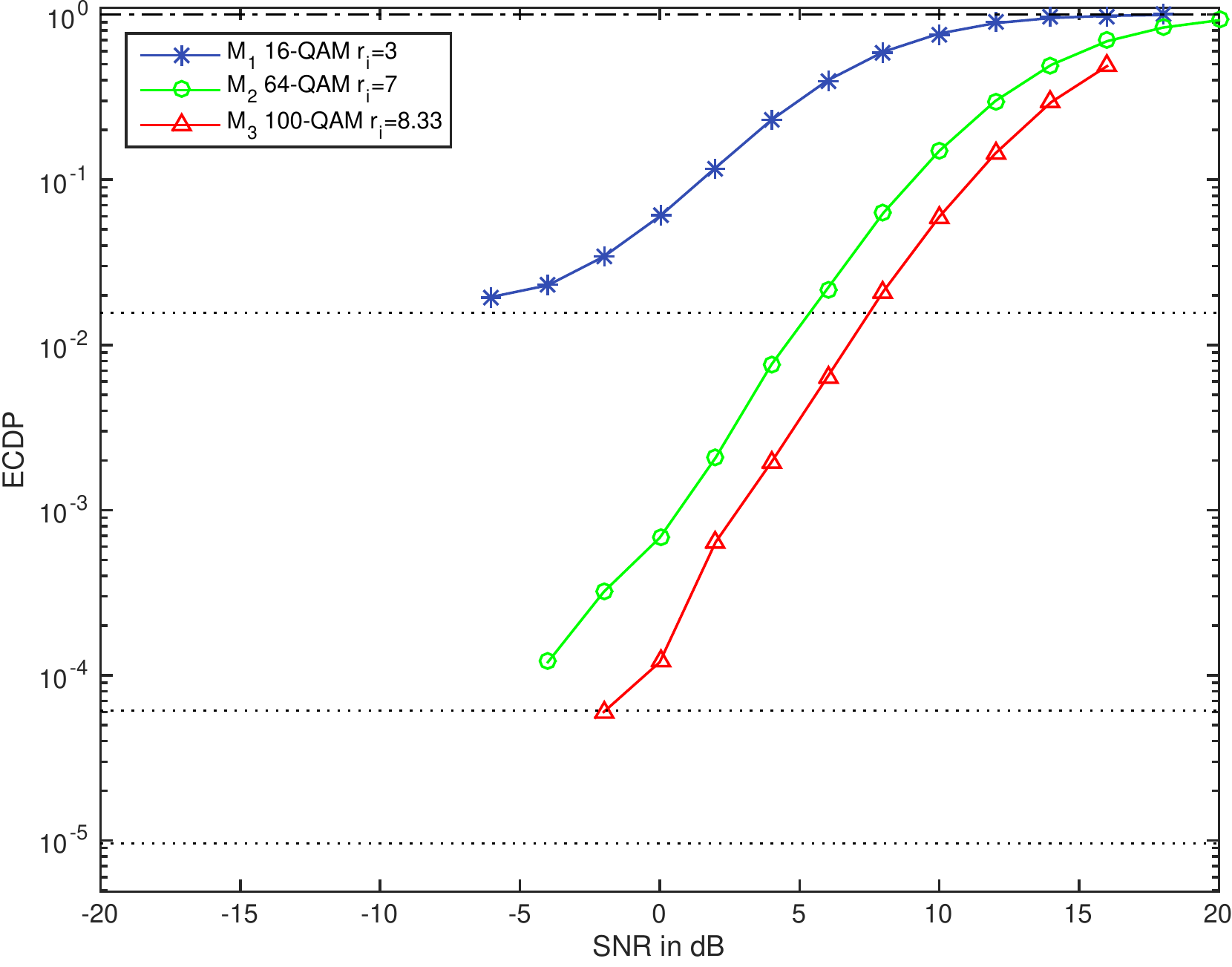}
	\caption{ECDP for 16-QAM, 64-QAM and 100-QAM Golden Code and $r_c \approx 5$}
	\label{Gold2_fig}
\end{figure}

\begin{table}[h!]
	\setlength{\tabcolsep}{5pt}
	\begin{center}
		\begin{tabular}{c r r r| r | r r r }
			\toprule
			& index	&	WR	&	$\lambda_1^2$ & Signaling Set & $r$ & $r_i$ & $r_c$\\ 
			\midrule
			
			$M_1$& $64$ & yes & 16 & 16-QAM & 8 & 3 & 5 \\
			$M_2$& $2^{14}$ & yes & 64 & 64-QAM & 12 & 7 & 5\\
			$M_3$& $103996$ & yes & 104 & 100-QAM & 13.29 & 8.33 & 4.96 \\
			\bottomrule\\
		\end{tabular}
		\caption{Sublattices of the Golden Code with $r_c \approx 5$}	\label{sublatGold2}
	\end{center}
\end{table}
In Figure \ref{Gold2_fig} we see that only the first lattice approaches its lower bound in the simulated region. It therefore seems that security is not only determined by $r_c$, but a reasonable balance between $r_i$ and $r_c$ has to be found.

\section{Conclusions and Future Work}
In this paper we have provided a practical design criterion for coset codes in fading MIMO wiretap channels. Simulations using the Alamouti and Golden code show that the proposed criterion indeed decreases the mutual information between the sender and the eavesdropper while maintaining equal reliability for a legitimate receiver in the high SNR regime.

Future work includes further investigations of well-rounded lattices, especially for higher indices. Also we plan on extending this work to other ST codes and perform an analysis of the effect of an outer code on the performance of the system.

\end{document}